\documentclass[12pt]{article}

\begin{document}
\title{Stable Solutions of the Double Compactified D=11 Supermembrane Dual}

\author{I. Mart\'{\i}n\footnote{isbeliam@fis.usb.ve}, J. Ovalle\footnote{jovalle@fis.usb.ve} and A. Restuccia\footnote{arestu@fis.usb.ve}\\
\vspace*{.25cm}\\
 Departamento de F\'{\i}sica, Universidad Sim\'on
Bol\'{\i}var, \\ Caracas, Venezuela.}
\date{}
\maketitle
\begin{abstract}The hamiltonian formulation of the supersymmetric closed 2-brane
 dual to the double compactified D=11 closed supermembrane is
presented. The formulation is in terms of two $U(1)$ vector fields
related by the area preserving constraint of the SUSY 2-brane.
Stable solutions of the field equations, which are local minima of
the hamiltonian, are found. In the semiclassical approximation
around the stable solutions the action becomes the reduction of
D=10 Super-Maxwell to the worldvolume. The solutions carry RR
charges as a type of magnetic charges associated with the
worldvolume vector field. The geometrical interpretation of the
solution in terms of $U(1)$ line bundles over the worldvolume is
obtained.
\end{abstract}

\newpage
The interest in D=11 supermembranes \cite{BST} has been renewed by
the realization that 11-dimensional supergravity may describe the
long distance behaviour of M-theory \cite{HT, PKT95, WIT, HW}. The
spectrum of the supermembrane in D=11 Minkowski target space has
been shown to be continuous when the $SU(N)$ regularization is
used \cite{dwln}. The analysis of the spectrum in the case of a
compactified target space has not been completed. Mean while, the
search for massless states, corresponding to the states of
11-dimensional supergra\-vity, still continues \cite{PKT95, PYI}.

In this work we analyze the closed $D=11$ supermembrane in a
target space $M_9$x$S^1$x$S^1$ \cite{mrt}. We consider the dual
formulation in terms of two vector fields. We perform the complete
canonical analysis, determine its physical hamiltonian and found
its minimal configurations. We show that the semiclassical
approximation of the action around the minimal configurations
describes the reduction of $D=10$ Super-Maxwell theory to the
world volume, showing that the configurations are stable ones. The
minimal configurations carry a type of magnetic charge that may be
interpreted as a RR charge which couples to the Kaluza-Klein
vector field.

 The 2-brane action in the form which includes an
independent auxiliary metric may be expressed as,
\begin{equation}
S(\gamma ,X)=-\frac{1}{2}\int d^{3}\xi \sqrt{-\gamma }\left(
\gamma ^{ij}\partial _{i}X^{\mu }\partial _{j}X^{\nu }\eta _{\mu
\nu }-1\right) \label{action}
\end{equation}
where $i$,$j$ denote the world volume indices while $\mu, \nu$
denote the 11 dimensional indices. $X^{\mu}$ define maps from the
3-dimensional world volume $M_3$, which will be assumed to be
$M_{3}=\Sigma$x$R$ where $\Sigma$ is a compact Riemann surface of
genus g, to the 11-dimensional space-time which is assumed to be
$M_{11-q}$x$\left[ S^{1}\right] ^{q}$ where $q$ is the number of
compactified coordinates. Each compactified coordinates $X$ is a
map from $\Sigma$x$R$ to $S^1$. It then satisfies
\begin{equation}
\oint_{c}dX=2\pi q_{c} \label{local}
\end{equation}
where $c$ is a basis of homology of dimension 1 elements, while
$q_c$ is an integral number associated to each element of the
basis. The set of $q_c$ for a basis of homology defines the
winding of the maps over $S^1$.

We will be interested in particular in the case $q\geq 2$ where the situation of irreducible winding may arise \cite{mrt}.

The dual formulation to (\ref{action}) may be obtained by considering
\begin{equation}
S=-\frac{1}{2}\int_{M_{3}}d^{3}\xi \sqrt{-\gamma }\left( \gamma
^{ij}L_{i}^{r}L_{j}^{s}\eta _{rs}+\gamma ^{ij}\partial
_{i}X^{m}\partial _{j}X^{n}\eta _{mn}-1\right)
-\int_{M_{3}}d^{3}\xi \partial _{j}A_{k}^{r}\epsilon
^{ijk}L_{i}^{r} \label{dualact}
\end{equation}
where $r$, $s = 1,...,\hat{q}$ and $m = 0,...,(10-\hat{q})$.
$A_{k}^{r}$ are $\hat{q}$ connection 1-forms which locally ensure
that $L^{r}_{i}$ satisfy the right constraints in order to recover
(\ref{action}), where $\hat{q}\leq q$, that is not all
compactified directions need be dualized. To implement the global
condition (\ref{local}) one has to impose the condition
\begin{equation}
\oint_{M_2}F(A^{r})=2\pi p^{r} \label{condition}
\end{equation}
where $M_2$ is a basis of homology of dimension 2 and summation
over all $p^r$ must be performed \cite{cmr}. The equivalence
between formulation (\ref{action}) and (\ref{dualact}) is ensured
provided summation in all winding is consider in (\ref{action})
and summation in all $p^r$ is taken in (\ref{dualact}) \cite{cmr},
\cite{wit2}; although this is strictly true for a compact
euclidean world volume.

The geometrical interpretation of (\ref{condition}) is given by
the Weil theorem \cite{weil} which ensures the existence of a U(1)
principle bundle over $M_3$ (compact, euclidean) and a connection
1-form over
 it whose curvature is identical to $F(A^r)$.

Funtional integration over $L^r$ leads to the action
\begin{equation}
S(\gamma ,X,A)=-\frac{1}{2}\int d^{3}\xi \sqrt{-\gamma }\left(
\gamma ^{ij}\partial _{i}X^{m}\partial _{j}X^{n}\eta
_{mn}+\frac{1}{2}\gamma ^{ij}\gamma
^{kl}F_{ik}^{r}F_{jl}^{r}-1\right) \label{2dbaction}
\end{equation}
The action (\ref{2dbaction}) for $q = 1$ and a gauge vector field corresponding to a trivial line bundle was first obtained in \cite{PKT95} following previous work in \cite{leigh} and \cite{smich}.

In order to analyze (\ref{2dbaction}), one has to distinguish the
case of an open membrane from the closed one. In the former, a
solution to the field equations arises by considering the time and
two spatial coordinates in the target space to coincide with the
local coordinates on the worldvolume. The semiclassical
approximation of the action around this solution yields the
reduction of D=10 Maxwell action to the worldvolume. The situation
for the case of a closed membrane is different. A solution where
the time and two spatial coordinates in the target space coincide
with the coordinates on the worldvolume requires three
compactified dimensions of the target, two of them identified with
the worldvolume spatial coordinates and the third one dua\-lized
to achieve a closed membrane with a U(1) gauge vector. There is
however another interesting scenario, which will be the point to
be analyzed in this paper, where $\hat{q} = q = 2.$ We will show
that there are stable solutions to the field equations arising
from the sector of the two U(1) gauge vectors, and not from the
metric on the worldvolume. The solutions are characterized by a
quantized type of magnetic charge in distinction to the above
discussed solutions. Furthermore the action around that solution
corresponds to the reduction of D = 10 Super-Maxwell to the
worldvolume. It is interesting to mention that in the case $q = 0,
1$ the solutions corresponding to the mi\-nima of the hamiltonian
are unstable. It is required at least two compactified directions,
$q = 2$, in order to have stable solutions \cite{mrt}.

If the metric $\gamma^{ij}$ is eliminated by using perturbative field equations, it was shown in \cite{PKT95} for $ \hat q = q = 1 $, that the Born-Infeld theory is obtained. See also \cite{arme}

To obtain the hamiltonian version of the theory we consider, in the usual way, the ADM decomposition of the metric
\begin{equation}
\begin{tabular}{l}
$\gamma _{ab}=\beta _{ab}$ \\
$\gamma ^{0a}=N^{a}N^{-2}$ \\
$\gamma ^{ab}=\beta ^{ab}-N^{a}N^{b}N^{-2}$ \\
$\gamma ^{00}=-N^{-2}$%
\end{tabular}
\label{ADM}
\end{equation}
where
\begin{equation}
\beta ^{ab}\beta _{bc}=\delta _{c}^{a}
\end{equation}
and
\begin{equation}
\sqrt{-\gamma }=N\sqrt{\beta }
\end{equation}
a,b are the 2 spatial indices.

The canonical action may then be rewritten as
\begin{equation}
S=\int_{M_{3}}d^{3}\xi \left( P_{m}\dot{X}^{m}+\Pi^{a}_r\dot{A}^{r}_{a}-{\cal H}%
\right)
\end{equation}
where the hamiltonian density may be expressed in the following
way
\begin{equation}
\begin{array}{l}
{\cal H} = \frac{N}{2}\left( \frac{P^{m}P_{m}}{\sqrt{\beta
}}+\frac{\beta _{ab} }{\sqrt{\beta }}\Pi _{r}^{a}\Pi
_{r}^{b}+\sqrt{\beta }\beta ^{ab}\partial _{a}X^{m}\partial
_{b}X_{m}-\sqrt{\beta }+\frac{1}{2}\sqrt{\beta }\beta ^{ac}\beta
^{bd}F_{ab}^{r}F_{cd}^{r}\right) \\
\\
+\Pi _{r}^{a}\partial _{a}A_{0}^{r}+N^{a}\left( \partial _{a}X^{m}P_{m}+\Pi^{b}_rF^{r}_{ab}\right)
\end{array}
\end{equation}
where $N$ and $N^{a}$ are Lagrange multipliers and $\beta_{ab}$ must satisfy
\begin{equation}
\beta _{ab}=\left( 1+\frac{1}{2}F^{r}F^{r}-\beta ^{-1}\Pi _{r}\Pi
_{r}\right) ^{-1}\left( \partial _{a}X^{m}\partial _{b}X_{m}+\beta
^{cd}F_{ac}^{r}F_{bd}^{r}-\beta ^{-1}\Pi _{ra}\Pi _{rb}\right)
\label{ecmovbeta}
\end{equation}
where $\beta$ is the determinant of the matrix $\beta_{ab}$. $F^rF^r$ and $\Pi_r\Pi_r$
are contracted with the metric $\beta_{ab}$.

We now consider the ligth-cone gauge fixing condition:
\begin{equation}
X^{+}=P_{0}^{+}{\cal T} \hspace{1cm} P^{+}=P^{+}_0\sqrt{W}
\end{equation}
where $ {\cal T} $ is the time coordinate on the world volume and $W$ the determinant of the metric over $ \Sigma $. Some authors take $W$ to be 1, however this condition cannot be imposed globally over any compact riemann surface. The metric over $\Sigma$ only appears in the action
through its determinant.

In the usual way one obtains
\begin{equation}
\partial _{a}X^{-}=\frac{-\partial _{a}X^{M}P_{M}-\Pi ^{b}_{r}F^{r}_{ab}}{P_{-}}
\label{eq:ffa}
\end{equation}
giving rise to the integrability constraint
\begin{equation}
\epsilon^{ab}\partial_b\left[\frac{\partial_aX^MP_M+\Pi^c_rF^r_{ac}}{\sqrt{W}}\right]=0
\label{intconstr}
\end{equation}
$M$ denote the transverse indices.

(\ref{intconstr}) is a necessary integrability condition for $X^{-}.$ In addition there are also global conditions on the right hand side member of (\ref{eq:ffa}) in order to have a well defined solution for $X^{-}$. They are
\begin{equation}
\oint_{c}\left(\frac{\partial_aX^MP_M+\Pi^b_{r}F^r_{ab}}{P_{-}}\right)d\xi^a=2 \pi n_c
\end{equation}
where $c$ is a basis of homology of dimension one. $P^{-}$ is obtained algebraically from  the constraint associated to the lagrange multiplier $N$.

The final form of the physical hamiltonian density is then
\begin{equation}
{\cal H}=\frac{1}{2}\frac{1}{\sqrt{W}}\left( P^{M}P_{M}+\beta +\frac{1}{2}%
\beta F^{r}F^{r}\right) -A_{0}^{r}\partial _{a}\Pi _{r}^{a}+\Lambda \epsilon
^{ab}\partial _{b}\left( \frac{\partial _{a}X^{M}P_{M}+\Pi
^{c}_rF^r_{ac}}{\sqrt{W}}\right) \label{hamiltoniano}
\end{equation}
where the first class constraint (\ref{intconstr}) has been left
as a restriction to the phase space variables. Its associated
gauge symmetry has not been fixed by the LC gauge condition, as
occurs in the canonical analysis for the membrane theory. The
first class constraint generates the algebra of the area
preserving diffeomorphisms in the extended phase space given by
$X, P $ and $A, \Pi$. The matrix $\beta_{ab} $ is an auxiliary
variable satisfying (\ref{ecmovbeta})

Any $2$x$2$ invertible matrix $ \beta_{ab}$ satisfies
\begin{equation}
\beta ^{cd}\epsilon _{ac}\epsilon _{bd}=\beta ^{-1}\beta _{ab}
\end{equation}
consequently (\ref{ecmovbeta}) may be simplify to
\begin{equation}
\beta _{ab}=\left( 1-\beta ^{-1}\Pi _{r}\Pi _{r}\right) ^{-1}\left( \partial
_{a}X^{M}\partial _{b}X_{M}-\beta ^{-1}\Pi _{ra}\Pi _{rb}\right)
\label{ecmovbeta2}
\end{equation}
that is $\beta_{ab}$ may be expressed as a function of $X^M$ and $ \Pi_{ra}$ only. We
now observe that the contraction
\begin{equation}
F^{r}F^{r}=F_{ab}^{r}F_{cd}^{r}\beta ^{ac}\beta ^{bd}
\end{equation}
may be rewritten as
\begin{equation}
F^{r}F^{r}=\frac{1}{2}\beta ^{-1}W(^{*}F^r)^{2} \label{eq:first}
\end{equation}
where
\begin{equation}
^{\ast }F^r=\frac{\epsilon ^{ab}}{\sqrt{W}}F^r_{ab}  \label{eq:second}
\end{equation}
The first parenthesis of the hamiltonian density in
(\ref{hamiltoniano}) may then be expressed as
\begin{equation}
\frac{1}{2}\frac{1}{\sqrt{W}}\left( P^{M}P_{M}+\beta \right) +\frac{%
1}{8}\sqrt{W}\left( ^{*}F^r\right) ^{2} \label{hamil2}
\end{equation}
showing that the terms $
\frac{1}{2}\frac{1}{\sqrt{W}}P_MP^M+\frac{1}{8}\sqrt{W}(^{\ast}F^r)^2$ formally
decouple from $ \frac{1}{2}\frac{1}{\sqrt{W}}\beta $ which depends on $X$ and $\Pi$
only. However $X^M$, $P_M$, $^*F$ and $\Pi_r$ are related by the area preserving
constraint.

We will now study the local minimal configurations of the
hamiltonian (\ref{hamiltoniano}). We will be in particular
interested in the stable local minima. To determine such
configurations we need to introduce a global geometric condition
in the phase space. The straightforward minima of the hamiltonian
arise over field configurations which are infinite dimensional and
are unstable, i.e. when $\Pi^a_r = 0$ we have infinite valleys
configurations as in the case of the supermembrane.

The minimal configurations are obtained for $\hat P$ and $\hat A$
satisfying
\begin{equation}
\hat P_{M}=0 \label{minconf}
\end{equation}

\begin{equation}
d^{*} {\hat F}^r=0 \label{cond}
\end{equation}
and
\begin{equation}
{\hat \Lambda} = 0
\end{equation}
(\ref{cond}) imply
\begin{equation}
^{\ast }{\hat F}^r=cte
\end{equation}
When the area of $\Sigma$ is normalized to $2 \pi$, the minimal
configuration are obtained for
\begin{equation}
^{\ast }{\hat F}^r=m^r
\end{equation}
where $m^r$ are integral numbers, characterizing the $U(1)$
principle bundle associated to each connection 1-forms $A^r$

We will consider the configurations for $ \Pi^{a}_{r}$, denoted by $ \hat{\Pi}^a_r,$
satisfying
\begin{equation}
\hat{\Pi}_{r}^{a}\hat{\Pi}_{s}^{b}\epsilon
_{ab}\epsilon^{rs}=n\sqrt{W} \hspace{1cm} n \neq 0 \label{zw}
\end{equation}
and the transverse coordinates
\begin{equation}
\hat{X}^{M}=0 \hspace{1cm} M=1,...,7.
\label{trans}
\end{equation}
For this particular configuration we obtain from
(\ref{ecmovbeta2}):
\begin{equation}
\hat{\beta}^{ab}=\left( 1-\hat{\beta}^{-1}\hat{\Pi}_{r}\hat{\Pi}_{r}\right)
^{-1}\left( -\hat{\beta}^{-1}\hat{\Pi}_{r}^{a}\hat{\Pi}_{r}^{b}\right)
\label{eq:se3}
\end{equation}
consequently
\begin{equation}
\hat{\beta}^{-1}\hat{\Pi}_{r}\hat{\Pi}_{r}=2  \label{eq:se4}
\end{equation}
and then
\begin{equation}
\hat{\beta}^{ab}=\hat{\beta}^{-1}\hat{\Pi}_{r}^{a}\hat{\Pi}_{r}^{b}
\end{equation}
where, using (\ref{zw})
\begin{equation}
\hat{\beta}=\det \left( \hat{\Pi}_{r}^{a}\hat{\Pi}_{r}^{b}\right) =\frac{1}{2%
}\hat{\Pi}_{r}^{a}\hat{\Pi}_{r}^{b}\hat{\Pi}_{s}^{c}\hat{\Pi}%
_{s}^{d}\epsilon _{ac}\epsilon _{bd} = \frac{n^2}{4}W
\label{eq:temporal}
\end{equation}

We will now consider the stability around these solutions. We take
\begin{eqnarray}
\Pi _{r}^{a}&=&\hat{\Pi}_{r}^{a}+\delta \Pi _{r}^{a}\\
X^{M}&=&\hat{X}^{M}+\delta X^{M} \nonumber
\end{eqnarray}
and expand $\beta$ up to second terms in $ \delta X^M$ and $ \delta\Pi^{a}_{r}$. We then obtain from (\ref{ecmovbeta2})
\begin{equation}
1-{\beta}^{-1}\Pi _{r}\Pi _{r}=-1+\partial X^{M}\partial
X_{M}=-1+\partial \delta X^{M}\partial \delta X_{M}
\end{equation}
which may then be substituted into (\ref{ecmovbeta2}) to get explicit expressions for the variations of $ \beta$. The terms linear in the variations are
\begin{equation}
2\left( \hat{\Pi}_{r}^{b}\hat{\Pi}_{s}^{d}\epsilon _{bd}\right)
\left( \delta \Pi _{r}^{a}\hat{\Pi}_{s}^{c}\epsilon _{ac}-\delta \Pi _{s}^{a}%
\hat{\Pi}_{r}^{c}\epsilon _{ac}\right)
\end{equation}
the first factor being proportional to $\sqrt{W}$ from (\ref{zw}).
The other factor may be expressed as a total derivative and
integrated out to give zero contribution. We conclude then that
(\ref{zw}), (\ref{trans}) is a stationary configuration of
(\ref{hamil2}).

The terms quadratic in the variations are
\begin{equation}
\beta = \hat{\beta} +
\hat{\beta}\left( \partial _{a}\delta X^{M}\partial _{b}\delta X_{M}\hat{%
\beta}^{ab}\right) +\frac{1}{2}\left( \delta \Pi
_{r}^{a}\hat{\Pi}_{s}^{c}\epsilon _{ac}-\delta \Pi
_{s}^{a}\hat{\Pi}_{r}^{c}\epsilon _{ac}\right) ^{2}+O(\delta ^{3})
\label{desbeta}
\end{equation}
showing that  (\ref{zw}), (\ref{trans}) is indeed a minimal configuration of the hamiltonian.

We have, because of the constraint on the momenta,
\begin{equation}
\delta\Pi^a_r = \epsilon^{ad}\partial_d\delta\Pi_r,
\end{equation}
then
\begin{equation}
\delta \Pi^a_r \hat{\Pi}^{c}_{s} \epsilon_{ac} \epsilon^{rs}=
\partial_c\delta\Pi_r.\hat\Pi^c_s.\epsilon^{rs}=
\epsilon^{cb}\partial_c \delta \Pi_r.\partial_b \hat{\Pi}_s \epsilon^{rs}
\end{equation}
It is convenient to define
\begin{equation}
{\cal A}_b = \delta\Pi_r\partial_b\hat\Pi_s\epsilon^{rs}
\label{zz}
\end{equation}

\begin{equation}
{\cal F}_{cb} = \partial_c{\cal A}_b-\partial_b{\cal A}_c,\hspace{2cm} ^{\ast }{\cal
F} = \frac{\epsilon^{cb}}{\sqrt{W}}{\cal F}_{cb}
\end{equation}
In terms of them (\ref{desbeta}) may be rewritten as
\begin{equation}
\beta = \hat{\beta} +%
\hat{\beta}\left( \partial _{a}\delta X^{M}\partial _{b}\delta X_{M}\hat{%
\beta}^{ab}\right) + \frac{W}{4}(^{\ast }{\cal F})^2+O(\delta
^{3}) \label{desbeta2}
\end{equation}
We then conclude that
\begin{equation}
\langle\frac{\beta}{\sqrt{W}}\rangle =
\langle\frac{\hat{\beta}}{\sqrt{W}} + O(\delta ^{3})\rangle
\label{desbeta3}
\end{equation}
if and only if
\begin{eqnarray}
\partial_a\delta X^M&=&0 \nonumber\\
{\cal F}_{cb}&=&0
\end{eqnarray}\label{ifonlyif}
i.e. the minimal configurations are strict up to closed 1-forms.
It is interesting that the interpretation of $ {\cal A}_b$ as a
1-form connection may be extended to consider $ {\hat \Pi}^{b}_{r}
{^{\ast}F^r} $ as its conjugate momenta.

In fact, the kinetic term in the action
\begin{equation}
\left\langle \Pi^a_r\dot A^r_a \right\rangle
\end{equation}
may be rewritten as
\begin{equation}
\left\langle \Pi^a_r\dot A^r_a \right\rangle = \left\langle \dot{{\cal A}}_b{\it \Pi^b} \right\rangle
\end{equation}
where
\begin{equation}
{\it \Pi}^b = \frac{1}{n}\hat \Pi^b_r{^{\ast}F^r}
\end{equation}

The area preserving generator, the first class constraint in (\ref{hamiltoniano}) may then be rewritten as
\begin{equation}
\partial_b{\it \Pi}^b + O(\delta ^{2}) = 0,
\end{equation}
This is, up to second order in the variations in the action, the
Gauss cons\-traint on the conjugate momentum to ${\cal A}$.

The hamiltonian (\ref{hamiltoniano}), in the semiclassical approximation, becomes then
\begin{eqnarray}
{\cal H} =&& \hat{\cal H}+\frac{1}{2}\frac{1}{\sqrt{W}}
\left(P_MP^M+\frac{n}{2}{\it \Pi}^a{\it \Pi}^bW_{ab}\right)+
\nonumber \\ &&\frac{1}{4}\sqrt{W} \left({\cal F}_{ab}{\cal
F}_{cd}W^{ac}W^{bd}+n\partial_a \delta X^M \partial_b \delta
X_MW^{ab}\right) \label{HSA}
\end{eqnarray}
where
\begin{equation}
W_{ab}=\frac{2}{n} \hat{ \beta}_{ab} \hspace{0.6cm}detW_{ab}=W
\label{metric}
\end{equation}
and
\begin{equation}
\hat {\cal H} = \frac{1}{8}\sqrt{W}\left(\left[m^r\right]^2+n^2\right)
\end{equation}
We thus conclude that the hamiltonian, in the semiclassical
approximation around the minimal configurations we found, exactly
describes the 10 dimensional Maxwell theory dimensionally reduced
to the world volume of the 2-brane. $X^M$ $M=1,...,7$ represent
the transverse coordinates to the 2-brane, while $ {\cal A}_b $
and ${\it \Pi}^b$ describe the Maxwell theory over the world
volume. This field content corresponds to a 2-Dbrane in 10
dimensions \cite{pol}, \cite{tay}. We notice that in order to
obtain this description the metric in the world volume is however
completely specified. It is given by (\ref{metric}) in terms of
the minimal configuration. The metric (\ref{metric}) is precisely
the canonical metric which naturally appears associated to the
monopole configurations \cite{mr}

This structure of the hamiltonian in the quadratic approximation
ensure that the local minima are stable solutions to the field
equations. In distinction to the case n = 0, in which the minimal
configurations expand an infinite dimensional subspace, as in the
D=11 Supermembrane theory over Minkowski space time.

Following (\ref{zz}) we may introduce
\begin{equation}
\hat{\cal A}_b = \frac{1}{2} \hat{\Pi}_r\partial_b
\hat{\Pi}_s\epsilon^{rs}
\end{equation}
and interpret (\ref{zw}) as
\begin{equation}
^{\ast}\hat{\cal F} = n
\label{ff}
\end{equation}

These solutions describe then monopole connections over the
Riemann surface $\Sigma$, generalizing the Dirac monopole over
$S^{2}$. The integer $n$ in (\ref{ff}) classifies all non-trivial
$U(1)$ line bundles over $\Sigma$. The general explicit
expre\-ssion for the gauge vectors was obtained in \cite{mr} in
terms of $U(1)$ connections 1-forms with non-trivial transitions
over $\Sigma$. See also \cite{fer}. A subset of these solutions
are dual to the ones obtained in \cite{mrt}, the interpretation is
now natural in terms of the $U(1)$ gauge vector field, even when
the global aspects of dualization are strictly valid on compact
worldvolumes.

The generalization of the above result to the super 2-brane in the
semiclassical aproximation arises in the same way. Instead of
(\ref{action}), we may consider the supermembrane action
\cite{BST}, and perform the double dualization procedure as in the
pure bosonic case.

The light cone gauge may be imposed by taking
\begin{eqnarray}
X^+&=& P^+_0{\cal T}\nonumber\\ P^+ &=& P^+_0\sqrt{W} \nonumber\\
\Gamma^+\psi &=& 0 \label{slcg}
\end{eqnarray}
We end up with the hamiltonian density
\begin{eqnarray}
{\cal H} =&&
\frac{1}{2}\frac{1}{\sqrt{W}}\left(P_MP^M+\beta+\frac{1}{2}\beta\beta^{ac}
\beta^{bd}{F}^r_{ab}{F}^r_{cd}\right)+\Pi^a_r\partial_aA^r_0+\Lambda\Phi
\nonumber \\
&&-\epsilon^{ba}\bar{\psi}\Gamma_-\Gamma_M\partial_a\psi\partial_bX^M
-\bar{\psi}\Gamma_{-}\Gamma^r\partial_b\psi\Pi^b_r
\label{superhamil2}
\end{eqnarray}
$\beta_{ab}$ satisfies the same constraint (\ref{ecmovbeta2}) and
the generator of the area preserving diffeomorphisms $\Phi$ has
the expression
\begin{equation}
\Phi =
\epsilon^{ab}\left(\partial_b\bar{\psi}\Gamma_-\partial_a\psi+\frac{1}{P^+_0}\partial_b
\left[\frac{\partial_aX^MP_M+F^r_{ac}\Pi^c_r}{\sqrt{W}}\right]\right)
\end{equation}
To quadratic order the only new contribution to ${\cal H}$ with
respect to (\ref{HSA}) is given by
\begin{equation}
-\bar{\psi}\Gamma_{-}\Gamma^r\partial_b\psi\hat\Pi^b_r
\end{equation}

The worldvolume canonical action arising from (\ref{superhamil2})
may be analyzed as in the bosonic case. The result is that the
configuration (\ref{zw}) together with $\psi = 0$, describe the
minimal configurations of (\ref{superhamil2}). The semiclassical
approximation around that minima describe the D=10 Maxwell
supermultiplet dimensionally reduced to three dimensions, in terms
of $X^M,\hspace{0.5cm} M=1,...,7$ scalar fields (the transverse
coordinates to the world volume of the 2D-brane), the world volume
vector potential ${\cal A}$ and eight $Sl(2,R)$ spinors arising
from the Majorana spinors $\psi$ satisfying the LCG condition
(\ref{slcg}).

The main property of the above minimal configurations is that they
des\-cribe isolated local minima (up to closed 1-forms over the
Riemann surface $\Sigma$) of the hamiltonian. That is, they are
locally stable solutions. This property distinguish the double
compactified closed D=11 supermembrane from the single
compactified closed one as discussed in \cite{mrt}.

The results in this paper extend then the construction obtained in
\cite{mrt} for the bosonic membrane to the supermembrane case. The
interpretation of the solution in terms of the $U(1)$ vector field
of the 2D-brane becomes then natural. In fact, the solution carry a
type of magnetic charge given by
\begin{equation}
Q = \int_\Sigma {\cal F} = \int d^2\xi\sqrt{W}. ^{\ast}{\cal F} =
Vol\Sigma.n
\end{equation}
The topological coupling of the $U(1)$ vector field on the
worldvolume to the RR 1-form connection $B$ of the IIA superstring
is naturally given by
\begin{equation}
\int_{M_3 = \Sigma x R} {\cal F}\wedge B
\end{equation}
We may then interpret Q as the RR charge carried by the closed
2D-brane.

In \cite{PKT95} it was argued how a classical closed membrane
configuration could be identified with a 0-brane which is needed
to describe the dynamics of the centre of mass motion of the
supermembrane. Moreover, it was argued that a closed membrane may
carry the 0-brane RR charge as a type of magnetic charge
associated with its world volume vector field, and that its centre
of mass motion should be described by the 0-brane $U(1)$
supersymmetric quantum gauge mechanics. This was interpreted as an
evidence that the 0-brane is included in the supermembrane
spectrum and hence that massless states are included in it.

 One further requirement should be added to the argument in
\cite{PKT95}. The solutions carrying the charge $Q$ should be
stable solutions of the membrane field equations. The solutions we
found have precisely the required stability property.

We have thus, constructed the physical hamiltonian of the
supersymme\-tric closed 2-brane dual to the double compactified
D=11 closed supermembrane with target space $M_9$x$S^1$x$S^1$. The
formulation is originally in terms of two $U(1)$ gauge fields over
the worldvolume, related by the area preser\-ving constraint. The
target space being $M_9$. We found the stable minimal
confi\-gurations of the hamiltonian and showed that the
semiclassical approxi\-mation of the action around those minimal
configurations correspond to the reduction of 10 dimensional Super
Maxwell to the worldvolume of the 2-brane. In that approximation,
the original formulation in terms of two gauge field is equivalent
to a theory in terms of only one $U(1)$ Maxwell potential, the
area preserving constraint being the Gauss constraint for it.

The minimal configurations are classified by three integer $m^r$,
$r=1,2$, and $n$. The integer $n$ was interpreted as a $RR$ charge
carried by the 2-brane and corresponds geometrically to the
integral number classifying the principle bundles over which the
$U(1)$ Maxwell connection is defined. When $m^r = 0$, $r=1,2$, we
then have from (\ref{eq:ffa}) and (\ref{trans})
\begin{equation}
\partial_a \hat{X}^-=0, \hspace{1.0cm} \partial_a \hat{X}^M=0
\end{equation}
that is, there is a solutions $\hat{X}^-=0$, $\hat{X}^M=0$. This
solution may be interpreted as a massless 0-brane moving in the
light cone of $M_9$ carrying a quantized RR charge. When $M^r \neq
0$, it follows from (\ref{eq:ffa}) that $\hat{X}^-$ is
proportional to $m^r\hat{\Pi}_r$. The target space time coordinate
$X^0$ becomes then a multivalued function over the worldvolume.

\end{document}